\documentclass[sigconf,letter,prologue,table,authorversion]{acmart} %EBG - Added prologue and table to avoid conflicts when loading xcolor for table
\pdfoutput=1

\usepackage{xcolor}
\usepackage{tabulary}
\usepackage{booktabs}
\usepackage{hyperref}
\usepackage{lipsum}
\usepackage{pifont}
\usepackage{enumitem}
\usepackage{xspace}
\usepackage{subcaption}
\newcommand{\filler}[1]{\relax}
\definecolor{OwnAzure}{HTML}{336699}
\definecolor{OwnCerulean}{HTML}{CAE2FE}
\definecolor{OwnOliveGreen}{HTML}{556B2F}
\definecolor{KamPurple}{HTML}{907C97}
\usepackage[colorinlistoftodos,prependcaption,textsize=tiny]{todonotes}% Add ,disable to the options, to hide the comments
\usepackage{xargs}                      % Use more than one optional parameter in a new commands
\newcommandx{\todolarge}[2][1=]{\todo[inline,size=\large,linecolor=OwnAzure,backgroundcolor=OwnCerulean,bordercolor=OwnAzure,#1]{#2}}
\newcommandx{\todoai}[2][1=]{\todo[inline,linecolor=OwnAzure,backgroundcolor=OwnCerulean,bordercolor=OwnAzure,#1]{#2}}

\newcommand{\customlabel}[1]{\label{#1}}
\newcounter{mainfinding}
\newcommand{\mainfind}[1]{\refstepcounter{mainfinding}\item[MF\themainfinding:\customlabel{mf:#1}]}
\newcommand{\mainfindref}[1]{MF\ref{mf:#1}} 
\newcounter{resquestion}
\newcommand{\resq}[1]{\refstepcounter{resquestion}\item[RQ\theresquestion:\customlabel{resq:#1}]}
\newcommand{\resqref}[1]{RQ\ref{resq:#1}} 

\newcommand\n{51}

\copyrightyear{2021}
\acmYear{2021}
\setcopyright{acmlicensed}\acmConference[ICPE '21 Companion]{Companion of the 2021 ACM/SPEC International Conference on Performance Engineering}{April 19--23, 2021}{Virtual Event, France}
\acmBooktitle{Companion of the 2021 ACM/SPEC International Conference on Performance Engineering (ICPE '21 Companion), April 19--23, 2021, Virtual Event, France}
\acmPrice{15.00}
\acmDOI{10.1145/3447545.3451197}
\acmISBN{978-1-4503-8331-8/21/04}

\begin{document}

% Camera Ready Instructions: Be sure to remove the headers from the top of both odd and even pages.
\fancyhead{}

\title{An Analysis of Distributed Systems Syllabi With a Focus on Performance-Related Topics}

\author{Cristina L. Abad}
\email{cabadr@espol.edu.ec}
\orcid{0000-0002-9263-673X}
\affiliation{%
  \institution{Escuela Superior Politecnica del Litoral}
  \city{Guayaquil}
  \country{Ecuador}
}

\author{Alexandru Iosup}
\email{a.iosup@vu.nl}
\orcid{0000-0001-8030-9398}
\affiliation{%
  \institution{VU Amsterdam}
  \city{Amsterdam}
  \country{the Netherlands}
}

\author{Edwin F. Boza}
\email{eboza@espol.edu.ec}
\affiliation{%
  \institution{Escuela Superior Politecnica del Litoral}
  \city{Guayaquil}
  \country{Ecuador}
}

\author{Eduardo~Ortiz~Holguin}
\email{leortiz@espol.edu.ec}
\affiliation{%
  \institution{Escuela Superior Politecnica del Litoral}
  \city{Guayaquil}
  \country{Ecuador}
}

\begin{abstract}
We analyze a dataset of \n~current (2019-2020) Distributed Systems syllabi from top Computer Science programs, focusing on finding the prevalence and context in which topics related to performance are being taught in these courses.
We also study the scale of the infrastructure mentioned in DS courses, from small client-server systems to cloud-scale, peer-to-peer, global-scale systems.
We make eight main findings, covering
    goals such as performance, and scalability and its variant elasticity;
    activities such as performance benchmarking and monitoring;
    eight selected performance-enhancing techniques (replication, caching, sharding, load balancing, scheduling, streaming, migrating, and offloading);
    and control issues such as trade-offs that include performance and performance variability.
\end{abstract}

\maketitle




\section{Introduction}
\label{sec:introduction}

At the start of the roaring (twenty-)twenties, higher education faces the important challenge of scaling to unprecedented numbers of students in Computer Science~(CS) and related fields.
But scaling is not the only important goal; equally, higher education must focus on how well the students acquire knowledge, skills, and experience relevant to the problems they will face in their professional life.
Performance is a relevant topic of general importance but with specific challenges in CS.
Though the WEPPE workshop has published various studies on the design and experience in teaching a standalone performance course~\cite{Persone:2017:WEPPE,Tay:2019:AnalyticalPerformanceModeling,Apte:2019:PerformanceAnalysisOfComputerSystemsAndNetworks}, a declining number of curricula include even a single course focusing entirely on performance~\cite{Persone:2017:WEPPE}.
Instead, performance aspects can appear in several courses in the curriculum, and especially in computer systems (e.g., ~\cite{Fisher:1992:PerformanceAnalysisParallelPrograms,Debray:2004:WritingEfficientPrograms,Ernst:2011:ManyCorePerformance}).\footnote{Furthermore, the Body of Knowledge of the IEEE/ACM 2013 CS Curriculum Guidelines~\cite{ACM:2013:Guidelines} lists topics and learning outcomes related to performance throughout several of its knowledge areas:
Algorithms and Complexity (AL),
Architecture and Organization (AR),
Computational Science (CN),
Information Assurance and Security (IAS),
Information Management (IM),
Intelligent Systems (NC),
Operating Systems (OS),
Platform-Based Development (PBD),
\textbf{Parallel and Distributed Computing (PD)},
Software Engineering (SE),
Systems Fundamentals (SF). (The emphasis on PD is ours.) %
} 
The scattered approach
hampers identifying common topics and sharing best teaching practices related to them. 
Addressing this problem, this work identifies and analyzes performance-related topics in a popular class of courses on Distributed Systems~(DS).

The choice of focus on DS courses is not arbitrary. 
First, DS courses appear often in curricula sanctioned by the IEEE/ACM guidelines~\cite{ACM:2013:Guidelines}. 
Other popular IEEE/ACM curricula, on Computer Engineering, Information Technology, and Software Engineering, also include important DS topics~\cite{ACM:2016:GuidelinesCE,ACM:2017:GuidelinesIT,ACM:2014:GuidelinesSE}.

Second, among the courses in which performance aspects can be discussed, Distributed Systems is a good target.
DS courses have a focus on the system view of the world, which is holistic by nature. Distributed systems present naturally a diverse set of levels and components where performance can be considered, e.g., from multi-node pools of execution threads to massive global-scale deployments, from processors to storage and to diverse sensors.
Last, distributed systems have a broad range of performance-aware applications, e.g., online gaming, video streaming, public transportation and high-performance scientific processing, so performance aspects can be explained with prominent and practical examples. 

We analyze \n~current Distributed Systems syllabi from top CS programs around the world and study whether or not they discuss issues related to performance in them.
Specifically, we 
seek to answer two main questions: \textit{(1) To what extent are performance-related topics being listed in the topic lists of these syllabi?} and \textit{(2) Do these syllabi include papers that have a strong focus on performance?}

Our work complements prior research in Computer Science education that has looked into performance engineering or teaching Distributed Systems (Section~\ref{sec:related}).
Closest to our work, is our previous study of these DS syllabi~\cite{Abad:2021:SIGCSE}, where we sought to answer four questions:
%(1) 
What are the most frequent DS course names?,
%(2) 
Which topics are commonly included in the syllabi?,
%(3) 
Which books and papers appear in the reading lists?,
%(4) 
and,
Do DS courses have a strong theoretical focus?
In contrast, this article focuses on the \textit{performance} aspects of DS courses. 

Our main contributions in addressing the main questions are:
\begin{enumerate}

    \item We define a set of research questions to analyze the presence of performance aspects in DS courses~(Section~\ref{sec:methods}).
    We cover common goals in performance engineering, such as
    improving performance and scalability, 
    common activities, such as benchmarking and monitoring, and
    common techniques, such as replication and scheduling.
    We also look at the scales of systems covered in DS courses, from small (e.g., cluster) to massive (e.g., datacenter) to global (e.g., Internet-scale).
    
    \item We present the main results of our data-driven analysis~(Section~\ref{sec:results}).
    We summarize a set of main findings, e.g., that, although DS courses cover often performance issues,  performance itself is not a topic commonly covered by these courses.
    For each question, we quantify the presence of specific performance issues among the course topics, and among the papers proposed as course reading material.
    
    \item We make a set of recommendations for the community teaching DS, aiming to improve the presence and diversity of performance topics~(Section~\ref{sec:discussion}).
    For example, we recommend that performance should be an explicit topic in DS courses.

\end{enumerate}

\section{Related Work}
\label{sec:related}
Prior historical and recent work on performance education within the CS curricula has looked into the design and experience in teaching courses that focus on performance, like
    Performance Evaluation~\cite{Shub:1989:PerformanceEvaluationCourse},
    Analytical Performance Modeling~\cite{Tay:2019:AnalyticalPerformanceModeling}, and
    Performance Analysis of Computer Systems and Networks~\cite{Apte:2019:PerformanceAnalysisOfComputerSystemsAndNetworks}.
For a comprehensive look at Performance Modeling (PM) courses, we refer the reader to the work of de Nitto Person\`{e}, who looked into ~75 courses on PM in the world and highlighted insights about their approach in a talk at WEPPE 2017~\cite{Persone:2017:WEPPE} and a later technical report~\cite{Persone:2020:Teaching}.

Issues about performance could also be added across the curricula to courses like
    Parallel Programming~\cite{Fisher:1992:PerformanceAnalysisParallelPrograms},
    a third-year programming course~\cite{Debray:2004:WritingEfficientPrograms}, and
    Computer Organization and Design~\cite{Ernst:2011:ManyCorePerformance}.
This paper complements these works by looking into the current practice of teaching Distributed Systems, identifying which issues related to performance are more commonly being taught.

The current work complements our prior study analyzing common patters in DS syllabi;
we refer the reader to~\cite{Abad:2021:SIGCSE} for a summary of the related work on education practices in teaching DS.

\section{Method for Analyzing Performance Aspects in DS Courses}
\label{sec:methods}

We propose a method for analyzing performance aspects in Distributed Systems (DS) courses. 
Our method is data-driven: starting from a previously collected dataset, we ask specific research questions.
For each question, we describe how to obtain its answer; this part of the method leverages the domain knowledge and expertise of the first two authors in teaching \textit{both} performance topics and distributed systems courses, a combined expertise of teaching such courses during over 30~academic years.

We analyze a dataset of DS syllabi that we manually collected and curated.\footnote{Available at: \url{https://doi.org/10.5281/zenodo.4290622}}
The dataset contains 51 syllabi from institutions in the top-100 of the 2019 Times Higher Education World University Rankings for Computer Science (CS).\footnote{Available at: \url{https://www.timeshighereducation.com/world-university-rankings/2019/subject-ranking/computer-science/}.}
To ensure that these syllabi are representative of current teaching practices, we selected syllabi from course offerings in 2019 and 2020.
When universities had more than one Distributed Systems course (e.g., Distributed Systems and Advanced Distributed Systems), we chose the most basic course (lower level); the reason for this is that the dataset was collected specifically with the intention of analyzing entry-level DS courses~\cite{Abad:2021:SIGCSE} 
and it is our view that performance issues should be part from the start in thinking about DS topics.
We manually curated the syllabi and built a table with thirteen fields: Rank, University, Country, Course name, Instructor, Course code, Semester/year, Links, Topic list, Textbook, other Recommended books, Papers listed as required reading, and Papers listed as optional or recommended reading.
We refer the reader to the related SIGCSE publication~\cite{Abad:2021:SIGCSE}\footnote{Available at: \url{https://arxiv.org/abs/2012.00552}} for more details on the collection and curation procedure.

The present study expands on our prior work by seeking to answer the following, detailed, research questions~(RQs):

\begin{description}

    \resq{goal-performance} \textbf{How frequently (and in what context) is the goal of \emph{performance} mentioned in the topic lists?}
We search for the string \emph{perf} in the topic lists and identify the number of syllabi in which it appears, as well as provide specific context in which it is mentioned in the topic lists;
for example Princeton's COS 418 Distributed Systems, lists the topic ``Reasoning about system performance'' in its syllabi.

    \resq{goal-scalability} \textbf{How frequently (and in what context) is the goal of \textit{scalability}, and its variant \textit{elasticity}, mentioned in the topic lists?}
We conduct a search similar to \resqref{goal-performance}, using the strings \emph{scale/scalable/scalability} for scalability and \emph{elast} for elasticity.

    \resq{activity-benchmarking} \textbf{How frequently (and in what context) is the activity of \textit{performance evaluation} mentioned in the topic lists?}
We search for the strings \emph{benchmark/test/eval} in the topic lists and analyze the results, seeking to determine if issues related to system performance benchmarking, testing, or evaluation are mentioned in the syllabi.

    \resq{activity-monitoring} \textbf{How frequently is the activity of \textit{performance monitoring} mentioned in the topic lists?}
    As for \resqref{activity-benchmarking}, we search for the string \emph{monitor}, which is applicable at various levels, from monitoring single variables to entire, globally distributed systems.
    
    \resq{technique-all} \textbf{How frequently (and in what context) are \textit{techniques used to improve performance} mentioned in the topic lists?}
Several performance-enhancing techniques can be discussed in the context of building distributed systems and their applications. Among them, we consider eight commonly used such techniques: replication, caching, sharding, load balancing, scheduling, streaming, migrating, and offloading.
Unless otherwise noted, we use primarily these terms to search for matching items in the topic lists. 
For the case of sharding, we also looked for the term \emph{partition} and consider as sharding if it refers to the same type of solution (e.g., data partitioning) but ignored it if it referred to network partitions (e.g., in the CAP theorem).
For scheduling, we considered both the term \emph{scheduling} and further \emph{queue/queueing} (the latter did not return any results);
we manually removed one instance related to transaction scheduling.
For streaming, we considered the terms \emph{stream/streaming} and also \emph{[message] queue} and \emph{publish[-subscribe]}.

    \resq{scale-all} \textbf{Which \textit{infrastructure scale} appears discussed in the DS course?}
    The scale of distributed systems ranges from a pair of computers acting in client-server roles, to massive global deployments of Internet-scale services. 
    We consider the scales (and \textit{terms}): 
    (i) \textit{client-server}; 
    (ii) \textit{cluster}, and the variant \textit{rack} (no hits); 
    (iii) \textit{datacenter}, with as variant spellings \emph{data center} and \emph{data[ ]centre};
    (iv) \textit{grid} and \textit{cloud} computing;
    (v) \textit{peer-to-peer} (\textit{p2p}), and the variant \textit{decentralized}; and
    (vi) \textit{global} scale, the variant \textit{Internet[-scale]} (for which we manually exclude references to Internet protocols), and \textit{Spanner}.
    
    \resq{control-all} \textbf{Which \textit{control} aspects appear in the DS course?}
    Systems engineers try to control various performance-related aspects, among which we focus on trade-offs (term \emph{trade}) and reducing performance variability (lists for terms \textit{variability/variable} and \textit{stability/stable/stabilize}, from which we eliminate topics that refer to functionality rather than performance, e.g., self-stabilizing algorithms).

    \resq{reading-all} \textbf{Which of the most commonly included academic papers mention performance or scale in their title?}
    More than half (29) of the syllabi in the dataset list academic papers as recommended or required readings.
    While we expect a high percentage of these papers to discuss issues related to performance in their evaluation sections,
    \resqref{reading-all} attempts to identify papers that focus very significantly on performance or scalability issues, such that these terms are explicitly mentioned in the paper titles.

\end{description}

\subsection{Threats to dataset validity}
\label{sec:threats}
The results in Section~\ref{sec:results} provide an analysis of the courses in our dataset;
we do not make broad claims about the overall state of the field.
Our analysis seeks to find trends and patterns within good syllabi, as indicated by their inclusion in ranked programs. This is a common approach in syllabi studies, e.g., see~\cite{Becker:2019:CS1Syllabi,Frechet:2020:Syllabi}.
However, any biases in the rankings may bias the results (e.g., by biasing towards research universities~\cite{Becker:2019:CS1Syllabi}).
Our data likely over-samples from anglophone countries (searches were in English), and from instructors who are comfortable sharing their syllabi.
The manual collection and curation approach has its limitations and we could have missed some courses, specially if they do not contain ``distributed'' in their name.
There may be reinforcement bias among the selected syllabi, as many syllabi are designed by consulting model curricula or are inspired from more established syllabi at other universities~\cite{Becker:2019:CS1Syllabi}.

\section{Results of Quantifying Performance Aspects in DS Courses}
\label{sec:results}

Herein we present the main results we have obtained in quantifying the presence of performance aspects in DS courses. 
Overall, our main observations are that: %our results (Section~\ref{sec:results}) show that:
\begin{description}
    \mainfind{goal-performance} 14\% DS courses mention \emph{performance} in the topic list.
    
    \mainfind{goal-scalability} 24\% mention issues related to \emph{scale} in the topic list.
    
    \mainfind{activity-benchmarking} None (0\%) list topics related to \textit{performance benchmarking} or \textit{evaluation} of distributed systems.
    
    \mainfind{activity-monitoring} A small fraction (6\%) of the topics list mention \textit{monitoring}.
    
    \mainfind{technique-all} Three-quarters of the performance-enhancing techniques we studied appear frequently in the topic lists, in order: {replication}~(61\%), {streaming}~(27\%), {caching}~(18\%), {scheduling}~(16\%), {sharding}~(8\%), and {load balancing}~(6\%). The other two, {migrating}~(only 2\%) and {offloading}~(zero), do not.
    
    \mainfind{scale-all} The topic lists include frequent explicit references to various infrastructure scales. Among them, the three most commonly mentioned are cloud computing~(29\%), Internet-scale~(27\%), and peer-to-peer~(22\%). No other scale appears above 8\%.
    
    \mainfind{control-all} Overall, there are few references to controlling performance issues.
    
    \mainfind{reading-all} The reading lists contain 8 and 34 different papers with \emph{performance} and \emph{scal[e/able/ability]} in their titles, respectively.
    
\end{description}

We next answer each of the research questions.

\subsection*{\normalsize \resqref{goal-performance}: How frequently (and in what context) is \emph{performance} mentioned in the topic lists?}
We find that 14\% of the syllabi contain the word \emph{performance} in the topic list~(\mainfindref{goal-performance}).
For context, Table~\ref{tab:terms:perf} lists the specific instances.

\rowcolors{2}{white}{gray!10}
\begin{table}[!t]
  \caption{Topics, in the topic lists of the surveyed syllabi, in which the string \emph{perf} appears.}
  \label{tab:terms:perf}
  \begin{tabular}{p{0.96\columnwidth}}
    \toprule
    \textbf{Topic} \\
    \midrule
    Reasoning about system performance \\
    Isolation and consistency semantics: Performance/usability tradeoffs \\
    Performance at scale \\
    Performance: eRPC \\
    Scalability vs. fault-tolerance vs. performance \\
    No compromises: Distributed transactions with consistency, availability, and performance (paper) \\
    NFS: Performance optimisations \\
    \bottomrule
\end{tabular}
\end{table}
\rowcolors{1}{white}{white}

\subsection*{\normalsize \resqref{goal-scalability}: How frequently (and in what context) is the goal of scalability, and its variant elasticity, included in the topics?}
We find that 24\% of the syllabi mention explicitly \emph{scal[e/able/ability]} in the topic list~(\mainfindref{goal-scalability}). We only find \emph{elasticity} mentioned in 2\% of the DS courses.
For context, Table~\ref{tab:terms:scal} lists the specific instances.

\rowcolors{2}{white}{gray!10}
\begin{table}[!t]
  \caption{Topics, in the topic lists of the surveyed syllabi, in which \emph{scal[e/able/ability]} or \emph{elast} appear.}
  \label{tab:terms:scal} \vspace{-6pt}
  \begin{tabular}{p{0.96\columnwidth}}
    \toprule
    \textbf{Topic} \\
    \midrule
    Don't settle for eventual: Scalable causal consistency for wide-area storage with COPS (paper) \\
    Scale-out key-value storage, Dynamo \\
    Case studies from industry: Google's Chubby fault-tolerant lock service, Google's Spanner scalable, fault-tolerant ACID database \\
    Large-scale data processing with MapReduce \\
    Performance at scale \\
    Large-scale data stores \\
    Load balancing: LARD, Internet-scale services \\
    Scalability issues and the concept of gossip \\
    Scalable services, reliability, and consistency: Scale and recovery for storage, leases, linearizable RPC for a replicated storage service \\
    Quality attributes~(availability/reliability, modifiability, scalability) \\
    Scalability vs. fault-tolerance vs. performance \\
    Scalability of blockchains \\
    Elastic services in the cloud: Managed services, mega-services and auto-scaling, request routing and load balancing: into the network, auto-sharding and sharded request routing \\
    \bottomrule
\end{tabular}
\end{table}
\rowcolors{1}{white}{white}

\subsection*{\normalsize \resqref{activity-benchmarking}: How frequently is the activity of performance evaluation mentioned?}
As finding \mainfindref{activity-benchmarking}, the term \emph{benchmark}, or the related strings \emph{test} and \emph{eval}, do not appear in any of the topic lists (0\%).

\subsection*{\normalsize \resqref{activity-monitoring}: How frequently (and in what context) is the activity of performance monitoring mentioned in the topic lists?}
We find that 6\% of the topic lists include monitoring~(\mainfindref{activity-monitoring}). We observe both low-level monitoring of variables, which is commonly used in distributed synchronization, and high-level monitoring of collections of servers.

\subsection*{\normalsize \resqref{technique-all}:  How frequently (and in what context) are techniques used to improve performance mentioned in the topic lists?}
We find that 6 out of 8 of the performance-enhancing techniques are well represented in the topic lists~(\mainfindref{technique-all}).
Among the performance-enhancing techniques we study, \textit{replication} is the most popular; this issue is listed in 61\% of the syllabi.
We note, however, that replication is frequently discussed only in the context of fault-tolerance and less as a performance-enhancing technique.

The second most popular technique is streaming, which appears in 27\% of the topic lists.
Topics dealing with caching, scheduling, sharding, and load balancing appear in 18\%, 14\%, 8\% and 6\% of the topic lists, respectively.
Migrating processes and offloading (parts of) applications appear less frequently as explicit topics, only 2\% and never, respectively.
Table~\ref{tab:terms:enhancing} lists the specific instances in which five of these techniques appear in the topic lists.

\rowcolors{2}{white}{gray!10}
\begin{table}[!t]
  \caption{Topics in the topic lists of the surveyed syllabi, in which performance-enhancing techniques (shown in bold) are mentioned. Of the 8~techniques we cover in this work, this table presents the results for replication, caching, scheduling, sharding, and load balancing. Topics that appear written in the same way in multiple topic lists appear only once in the table, with multiple instances denoted in parentheses.}
  \label{tab:terms:enhancing}
  \small
  \begin{tabular}{p{0.96\columnwidth}}
    \toprule
    \textbf{Topic} \\
    \midrule %Replication
    Caching and \textbf{replication} \\
    Consensus: Passive \textbf{replication} (raft), active \textbf{replication}, lazy \textbf{replication} (gossip, Bayou) \\
    Consistency and \textbf{replication} (x6) \\
    Data \textbf{replication} \\
    Data \textbf{replication} and consistency: Overview, consistency models \\
    Distributed \textbf{replication} \\
    Fault-tolerance: \textbf{Replicated} and fusible state machine approaches \\
    MapReduce and \textbf{replicated} state machines (x2) \\
    Primary-backup \textbf{replication} (x3) \\
    \textbf{Replicated} state machines (x2) \\
    \textbf{Replication} (x4) \\
    \textbf{Replication} (primary/backup, quorum protocols, sequential and causal consistency, client-centric models) \\
    \textbf{Replication} (\textbf{replicated} state machine, primary/backup, quorum \textbf{replication}) \\
    \textbf{Replication} control  \\
    \textbf{Replication} in distributed systems \\
    \textbf{Replication} in the HARP file system \\
    \textbf{Replication}, active \textbf{replication}, primary-backup \textbf{replication}, gossip-based \textbf{replication} \\
    \textbf{Replication}, data-centric consistency \\
    \textbf{Replication}: Epidemic algorithms, Bayou \\
    Scalable services, reliability, and consistency: Scale and recovery for storage, leases, linearizable RPC for a \textbf{replicated} storage service \\
    \midrule %Caching (9)
    \textbf{Cache} consistency (x2) \\
    \textbf{Caching} and consistency: NFS and the Web \\
    \textbf{Caching} and replication \\
    Consistency and replication: Consistency models, consistency protocols, replica management, \textbf{caching} \\
    Distributed file systems: Architecture, \textbf{caching}, semantics \\
    Facebook photo \textbf{cache} \\
    \textbf{Memcache} \\
    Web \textbf{caching} and consistent hashing \\
    \midrule %Scheduling (7)
    Distributed Computing: \textbf{Scheduling} and Resource Management \\
    Local \textbf{Scheduling}: Scale-out threads \\
    Multiprocessor and Distributed \textbf{Scheduling} \\
    \textbf{Scheduling} (x3) \\
    \textbf{Scheduling} and Load Balancing in distributed systems \\
    \midrule %Sharding
    \textbf{Sharding} \\
    \textbf{Sharding} and distributed transactions \\
    Elastic services in the cloud: [...] auto-\textbf{sharding} and \textbf{sharded} request routing (see full topic name in Table~\ref{tab:terms:scal}) \\
    Data partitioning strategies (authors' note: same as sharding) \\
    \midrule %Load balancing
    \textbf{Load balancing}: LARD, internet-scale services \\
    Scheduling and \textbf{load balancing} in distributed systems \\
    Elastic services in the cloud: [...] request routing and \textbf{load balancing}: into the network, [...]  (see full topic name in Table~\ref{tab:terms:scal}) \\
    \bottomrule
\end{tabular}
\end{table}
\rowcolors{1}{white}{white}

\subsection*{\normalsize \resqref{scale-all}: Which infrastructure scale is discussed?}
We find that the topic lists include frequent explicit references to various infrastructure scales~(\mainfindref{scale-all}).
In increasing order of the maximum scale observed in practice,\footnote{Whereas the order server, cluster, datacenter is natural, the overuse of terms grid, cloud, and peer-to-peer makes their relative ordering more difficult to consider. We propose an ordering based on number of geographical locations. The largest grids span tens to hundreds of locations; public clouds span at most a few tens of geographically-distributed datacenters. In contrast, the largest P2P systems, for example, the large-scale deployments of Bittorrent for file-sharing~\cite{DBLP:conf/hpdc/WojciechowskiCPI10}, and of BOINC or similar for volunteer computing~\cite{DBLP:journals/jpdc/IglesiasKM12}, reached at-peak hundreds of millions of locations in the world.} the frequencies are:
(i)~\textit{client-server}, only 2\%; 
(ii)~\textit{cluster}, 6\%, with the variant \textit{rack} receiving zero explicit mentions; 
(iii)~\textit{datacenter}, 8\%;
(iv)~\textit{grid} and \textit{cloud} computing, 4\% and 29\%, respectively, indicating their popularity;
(v)~\textit{peer-to-peer}, 24\%, including the variant \textit{decentralized}, at 1\%; and
(vi)~\textit{global} scale, 27\%, including the variant \textit{Internet[-scale]}, at 4\%, and \textit{Spanner}, at 22\%, the most common example of a global-scale system.

The very common presence of cloud computing is not surprising, given the wide applicability of the concept and its popularity in current practice. In contrast, Spanner~\cite{DBLP:journals/tocs/CorbettDEFFFGGHHHKKLLMMNQRRSSTWW13} is surprisingly popular; we attribute this to the high-visibility publication venue, the brand of the Google engineering team, the high quality of concepts present in the Spanner article, and also to the maturity of both design and implementation that the team explains is the outcome of over five years of system, software, and performance engineering.

\subsection*{\normalsize \resqref{control-all}: Which control aspects appear in the DS course?}

We find few references to controlling performance issues~(\mainfindref{control-all}).

The systems view is often focused on trade-offs. Performance is a common element in trade-offs in practice, against functional system properties such as consistency and non-functionals such as availability (e.g., in the PACELCA view extending the CAP theorem in the last decade), but also against other non-functionals, such as energy consumption and cost. 
We thus find it surprising that the term trade-off appears in only 6\% of the topic lists. 

Given the prominence of cloud among the topic lists, and the prevalence of the phenomenon of performance variability in clouds in the past decade~\cite{DBLP:conf/ccgrid/IosupYE11,DBLP:conf/nsdi/UtaCDJRMRI20}, we find it even more surprising that zero~(0\%) of the topic lists include \textit{variability} or related terms.

\subsection*{\normalsize \resqref{reading-all}:  Which of the most commonly included academic papers mention performance or scale in their title?}
The reading lists contain 8 and 34 different papers with \emph{performance} and \emph{scal[e/able/ability]} in their titles, respectively~(\mainfindref{reading-all}).
Table~\ref{tab:papers} contains the list of paper titles that contain these terms.

\rowcolors{2}{white}{gray!15}
\begin{table}[thb]
  \caption{List of paper titles included in the DS syllabi as required or recommended readings, that contain performance (top, 8 papers) or scal(e/able/ability) (bottom, 34 papers).}
  \label{tab:papers}
  \small 
  \begin{tabular}{p{0.96\columnwidth}}
    \toprule
    \textbf{Paper Title} \\
    \midrule %Performance
        Building Secure High-Performance Web Services with OKWS \\
        Characteristics of Scalability and Their Impact on Performance \\
        Implementation and Performance of Munin \\
        No compromises: Distributed transactions with consistency, availability, and performance \\
        Paxos Replicated State Machines as the Basis of a High-Performance Data Store \\
        Performance debugging for distributed systems of black boxes \\
        The Akamai Network: A Platform for High-Performance Internet Applications \\
        The Performance Implications of Thread Management Alternatives for Shared-Memory Multiprocessors \\    
    \midrule %Scal*
        A scalable content-addressable network \\
        Algorand: Scaling Byzantine Agreements for Cryptocurrencies \\
        Bitcoin-NG: A Scalable Blockchain Protocol \\
        Characteristics of Scalability and Their Impact on Performance \\
        Chord: a scalable peer-to-peer lookup protocol for internet applications \\
        Cluster-based file replication in large-scale distributed systems \\
        Dapper, a Large-Scale Distributed Systems Tracing Infrastructure \\
        Discretized Streams: A Fault-Tolerant Model for Scalable Stream Processing \\
        Don't Settle for Eventual: Scalable Causal Consistency for Wide-Area Storage with COPS \\
        Experiences with a Distributed, Scalable, Methodological File System: AnalogicFS \\
        Exploiting a Natural Network Effect for Scalable, Fine-grained Clock Synchronization \\
        Frangipani: A Scalable Distributed File System \\
        Highly Scalable Algorithm For Distributed Real-Time Text Indexing \\
        Implementing linearizability at large scale and low latency \\
        Large-scale cluster management at Google with Borg \\
        Lessons from Giant-Scale Services \\
        OceanStore: An Architecture for Global-Scale Persistent Storage \\
        Omega: Flexible, scalable schedulers for large compute clusters \\
        On scalable and efficient distributed failure detectors \\
        PRAM: A Scalable Shared Memory \\
        Pastry: Scalable, Decentralized Object Location, and Routing for Large-Scale Peer-to-Peer Systems \\
        Pregel: A system for large-scale graph processing \\
        SVE: Distributed Video Processing at Facebook Scale \\
        SWIM: Scalable Weakly-consistent Infection-style Process Group Membership Protocol \\
        Scalable Application Layer Multicast \\
        Scaling Distributed Machine Learning with the Parameter Server \\
        Scaling Memcache at Facebook \\
        Sinfonia: A new paradigm for building scalable distributed systems \\
        Spotify - Large Scale, Low Latency, P2P Music-on-Demand Streaming \\
        Storage management and caching in PAST, a large-scale, persistent peer-to-peer storage utility \\
        Tapestry: A Resilient Global-scale Overlay for Service Deployment \\
        TensorFlow: A System for Large-Scale Machine Learning \\
        The tail at scale \\
        ZooKeeper: Wait-free Coordination for Internet-scale Systems \\
    \bottomrule
\end{tabular}
\end{table}
\rowcolors{1}{white}{white}

%\section{Recommendations on Enhancing the Performance Component of DS Courses}
\section{Recommendations on How DS Syllabi Address Performance}
\label{sec:discussion}

\textbf{Include more performance topics, and more explicitly, in the DS curriculum.}
Less than one fourth of the syllabi explicitly mention \emph{performance} (14\%) or \emph{scale} (24\%) in the topic lists.
However, these issues may be well covered within other topics in the topic lists (e.g., remote invocation and cluster management, which appear in 67\% and 24\% of the topic lists~\cite{Abad:2021:SIGCSE}) or at independent performance-focused courses at the corresponding institutions~\cite{Persone:2017:WEPPE} (for example, the Politecnico di Milano which has a DS course and a course on Performance Evaluation of Computer Systems~\cite{Persone:2020:Teaching}).

\textbf{Focus on performance \& scale as first-class concepts.}
The software engineering community has repeatedly noted that performance and scalability issues tend to be an afterthought in the development process~\cite{Menasce:2004:Performance,Liu:2011:software,Ousterhout:2017:Performance}.
These issues should be explicitly listed in the syllabi and not buried under other topics.
Doing so would increase their visibility, solidify the idea that software performance is important, and possibly even help attract more students to an important course that is frequently listed as elective.

\textbf{Add learning goals focusing on benchmarking or experimental evaluation of DS.}
None (0\%) of the syllabi in our dataset list topics related to benchmarking or experimental evaluation of distributed systems.
Given that courses focusing on this issue are even less popular than Distributed Systems courses, and given that important curricular initiatives~\cite{ACM:2013:Guidelines} have found this to be an important topic to be included in CS programs, we think the community should consider making distributed systems evaluation a first-class topic instead of relegating it to a project or ignoring it altogether.

\textbf{Consider including monitoring in DS syllabi.}
Another essential activity in distributed systems, monitoring, appears to be under-valued by current DS syllabi.
We recommend introducing this topic, which raises important conceptual challenges and has resulted in notable practical results such as CERN's MonALISA~\cite{legrand2009monalisa}, Ganglia~\cite{massie2004ganglia}, and more recently, DevOps pipelines~\cite{Miglierina:2017:Omnia} leveraging  Prometheus and Grafana.\footnote{See~\url{https://grafana.com/oss/prometheus/} and \url{https://grafana.com/oss/grafana/}}
While monitoring issues may be discussed in a Software Engineering course, the challenges of monitoring large-scale systems are arguably better understood in a DS class.
Furthermore, teaching performance-enhancing techniques without also including how to properly verify how they work (i.e., through testing, analysis, monitoring) greatly reduces the value and impact of the technique.\footnote{We thank our anonymous referees for pointing this out.}

\textbf{Increase coverage of performance-enhancing techniques used in practice and their trade-offs.}
Many performance-\\enhancing techniques are already widely present in DS syllabi, starting with replication.
However, we found that important techniques that are commonly used in practice, such as process migration and component offloading, have a negligible presence in the topic lists of modern Distributed Systems courses. 
Also neglected are aspects critical to controlling the performance of distributed systems, such as understanding the trade-offs where performance plays a prominent role and reducing performance variability, do not currently appear explicitly in the DS syllabi we study.
While these trade-offs are likely discussed in class, explicitly listing the trade-offs as topics being studied could help students be more aware of their importance and can also help highlight the relevance of the DS course in engineering practice.

\section{Conclusion}
\label{sec:conclusion}

We identified the need to study the presence of performance aspects in Distributed Systems syllabi. 
Addressing this need, we conducted an analysis of \n~Ditributed Systems syllabi, focusing on finding the prevalence and context in which topics related to performance are being taught in these courses.
Our analysis represents the most comprehensive study of its kind, to date. 
It covers for performance engineering a number of 
common goals, activities, techniques, and control aspects. 
We also study the scale of the infrastructure mentioned in DS courses, from small client-server systems to cloud-scale, peer-to-peer, global-scale systems.
We make eight main findings, covering 
goals such as performance, and scalability and its variant elasticity;
activities such as performance benchmarking and monitoring;
eight selected performance-enhancing techniques, that is, replication, caching, sharding, load balancing, scheduling, streaming, migrating, and offloading;
and control issues such as trade-offs that include performance and performance variability.

We envision the use of this work in the development of DS courses that focus more on performance topics. We also foresee our method for syllabus analysis being leveraged by directors of education in full-curriculum analysis, as element in their reporting and support in their decision-making processes.

\begin{acks}
We thank the anonymous WEPPE reviewers that thoroughly read our paper and provided invaluable feedback that helped improve the quality of this manuscript.
Part of this work was sponsored by a Google Faculty Research Award (CA, EB), and the NWO (AI), projects MagnaData and OffSense.
\end{acks}

\bibliographystyle{ACM-Reference-Format}
\bibliography{references}

\end{document}